\documentclass[twocolumn,superscriptaddress,longbibliography,aps,pra,preprintnumbers]{revtex4-1}
\usepackage{graphicx}
\usepackage{bm}
\usepackage{color}
\usepackage{epstopdf}
\usepackage{amsmath}
\usepackage{amssymb}
\usepackage{ulem}
\usepackage{xfrac}
\usepackage[urlcolor=blue,colorlinks=true,citecolor=blue,linkcolor=blue,pdfstartview={FitH},bookmarks=false]{hyperref}

\newcommand{\AP}[1]{\textcolor{black}{#1}}
\newcommand{\AK}[1]{\textcolor{black}{#1}}

\newcommand{\Imag}{\text{Im}}
\graphicspath{{figures/}{./figures/}{.}}
\begin{document}


\title{Majorana Bound State leakage to impurity in Su--Schrieffer--Heeger--Rashba scenario}

\author{Aksel Kobia\l{}ka}
\email[e-mail: ]{akob@kft.umcs.lublin.pl}
\affiliation{Institute of Physics, M. Curie-Sk\l{}odowska University, 
			pl. M. Sk\l{}odowskiej-Curie 1, 20-031 Lublin, Poland}

\author{Andrzej Ptok}
\email[e-mail: ]{aptok@mmj.pl}
\affiliation{Institute of Nuclear Physics, Polish Academy of Sciences, 
			ul. W. E. Radzikowskiego 152, 31-342, Krak\'ow, Poland.}

\date{\today}

\begin{abstract}
\AK{We show anomalous features of Majorana Bound State leakage in the situation where topological Rashba nanowire is dimerized according to the Su--Schrieffer--Heeger (SSH) scenario and an impurity is present at one of the ends of the system.
We find that two topological branches: usual, indigenous to Rashba nanowire and dimerized one, existing as a result of SSH dimerization of nanowire, have different asymmetry of spin polarization that can be explained by opposite order of bands taking part in topological transitions. 
Additionally, introduction of an impurity to the dimerized nanowire influences the leakage of Majorana bound states into the trivial impurity, due to the emergence of Andreev bound states that behave differently whether the system is or is not in topological phase.
This results in the pinning of zero energy states to the impurity site for some range of parameters.}
\end{abstract}

\maketitle

\section{Introduction}
\label{sec.intro}

\AK{Systems exhibiting an existence of Majorana Bound States (MBS) show a great promise for the emergence of a new branch of quantum computing --- a topological quantum computing, relying on topological superconductors.
Quantum computing is a steadily growing field of both physics and nanotechnology, however a working example of its topological counterpart is still yet to be presented. 
A presumed advantage of topological quantum computing over a "regular" one, is the property of fault--tolerant computing~\cite{aasen.hell.16}.}
In order to achieve this, non--Abelian quasiparticles~\cite{kitaev.01} have to be employed, hence the interest in MBS, which are believed to possess such properties~\cite{nayak.simon.08}.

Recently, such \AP{\it quasi}particles have been experimentally uncovered in numerous examples, both in one dimensional (1D) systems (e.g., in form of zero--energy bound states localized at the ends of nanowires deposited upon a surface, due to interplay between spin--orbit coupling, superconductivity and magnetic field)~\cite{mourik.zuo.12,das.ronen.12,nadjperge.drozdov.14,pawlak.kisiel.16,deng.vaitiekenas.16,nichele.drachmann.17,jeon.xie.17,kim.palaciomorales.18,lutchyn.bakkers.18,gul.zhang.18,Fornieri-2019,Ren-2019} or two dimensional systems (2D) (e.g., edge states around an superconducting island)~\cite{menard.17,menard.mesaros.18,palacio.mascot.18}.

Dimerization alone can allow for topological transition, even if superconductor is not present in the system~\cite{Su_1979}.
For instant, in the Su--Schrieffer--Heeger (SSH) model~\cite{Su_1979,Heeger_1988}, two different bonds between atoms are assumed,  which makes the atoms dimerize due to Peierls instability.
This phenomena generated some interest, but mainly using the combination of Kitaev~\cite{kitaev.01} and SSH models~\cite{Wakatsuki_2014,Wang_2017,Ezawa_2017,Chitov_2018,Yu_2019,Hua_2019}.
Therefore, we combine aforementioned SSH dimerisation with Rashba nanowire properties in order to obtain a Su--Schrieffer--Heeger--Rashba (SSHR) model.

\begin{figure}[!h]
\includegraphics[width=0.9\linewidth,keepaspectratio]{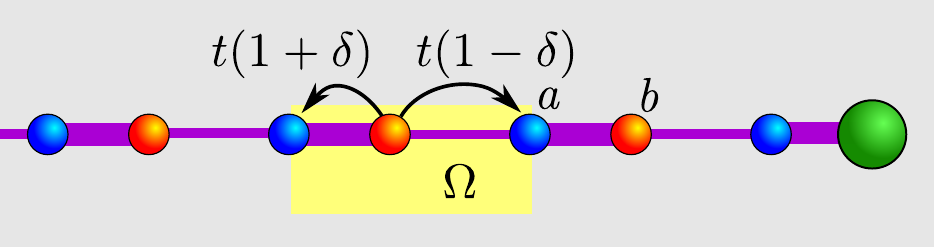}
\caption{Schematics of the dimerized nanowire proximitized to the isotropic superconductor. 
Modulation of the hopping integral $\delta$ corresponds to the shifts in positions between the neighboring $a$ and $b$ sites in the \textit{unit cell} $\Omega$ (marked by the yellow frame). 
\AK{Sites on selected sublattices $a$ and $b$ are marked by blue and red colors, respectively, while green corresponds to an additional impurity site connected to end of the nanowire.}
}
\label{fig.scheme}
\end{figure}

MBS as an edge phenomena, tends to leak to the furthest elements of the system, even if those parts (e.g. impurity) do not manifest any topologically non--trivial nature~\cite{ptok.kobialka.17,kobialka.ptok.18.ring,kobialka.ptok.18.acta,kobialka.ptok.18.plaq}.
Here, we check how the leakage of MBS behaves when impurity is attached to the end of dimerized Rashba nanowire, within the SSH scenario (cf.~Fig.~\ref{fig.scheme}), depending on the order of the bond strength and thus the type of the bond between last two sites in the system.

The paper is organized as follows.
This section is an introduction to the paper.
In Sec.~\ref{sec.model} we introduced the SSH model of the dimerized Rashba nanowire and methods.
In Sec.~\ref{sec.results} we discuss results obtained by numerical calculation.  
Then, we summarize our results in Sec.~\ref{sec.sum}.

\section{Methodology}
\label{sec.model}

We consider a SSH analogue of Rashba nanowire, where the 1D semiconducting nanowire that is deposited on superconducting substrate (Fig.~\ref{fig.scheme}) is modified with an alternating order of the weak and strong bonds (or \textit{vice versa}) that emulate and SSH scenario.
We model the distance between the neighboring sites $a$, $b$ (forming the \textit{unit cell} ${\Omega}$) by modulation of the hopping integral $\delta$ that effectively changes the probability of electron transport between neighboring sites. 
Similar modulations affect also the spin-orbit Rashba interactions (Eq.~\ref{Rashba_term}).
In  a natural way, the SSH model describes a system with two sublattices (sites $a$- and $b$-type).

\paragraph*{Microscopic model. --} 
Our system can be described by the Hamiltonian $\mathcal{H} = \mathcal{H}_{0}+\mathcal{H}_\text{so} + 
\mathcal{H}_\text{prox}$.
First term:
\begin{eqnarray}
\mathcal{H}_{0} = &-& \sum_{i \sigma} \left[ t(1+ \delta)c_{i a \sigma}^{\dagger} c_{i b \sigma}  
+ t(1- \delta) c_{i a\sigma}^{\dagger} c_{i-1 b \sigma} +\mbox{\rm h.c.} \right]  \nonumber \\
&-& \sum_{s \in \Omega} \sum_{i,\sigma}  \left( \mu + \sigma h \right) 
c_{is\sigma}^{\dagger} c_{is\sigma},
\label{eq.SSH_part}
\end{eqnarray}
describes a SSH-like nanowire.
The operators $c_{is\sigma}^{\dagger}$ ($c_{is\sigma}$) denotes creation (annihilation) of the electron with spin $\sigma$ in {\it i}-th \textit{unit cell} and sublattice $s$ (e.g., site $a$ or $b$). 
$\mu$ is the chemical potential and $h$ denotes the magnetic field in the Zeeman form. 
$(1 \pm \delta)$ is a periodic variation of hopping integral $t$ between nearest neighboring sites, i.e., between sites in different sublattices.
We assume, similar modulation also for the spin-orbit Rashba term:
\begin{eqnarray}
\mathcal{H}_\text{so} = &-& i \sum_{i \sigma \sigma'} \left[ \lambda(1+ \delta) c_{i a \sigma}^{\dagger} 
( \sigma_{y} )_{\sigma\sigma'} c_{i b \sigma'} \right. \label{Rashba_term} \\ \nonumber 
&+& \left. \lambda(1- \delta)c_{i a \sigma}^{\dagger} ( \sigma_{y} )_{\sigma\sigma'} 
c_{i-1 b \sigma'} \right] + \mbox{\rm h.c.} 
\end{eqnarray}
where $\sigma_{y}$ is the second Pauli matrix and $\lambda$ describes the strength of the spin-orbit coupling.
Last term models the BCS-like superconducting gap, that arises from proximity effect, i.e., deposition of nanowire on superconducting surface~\cite{chang.albrecht.15}:
\begin{eqnarray}
\mathcal{H}_\text{prox} =   \sum_{is} \left( \Delta 
c_{is\uparrow}^{\dagger} c_{is\downarrow}^{\dagger}  + 
\Delta^{\ast} c_{is\downarrow}c_{is\uparrow} \right) .
\end{eqnarray}
Impurity is treated as a additional site connected to the nanowire that is not affected by proximity effect, i.e. $\Delta_\text{imp} = 0$.

In typical situation of homogeneous nanowire, transition from the trivial to non--trivial topological phase occurs for some critical value of magnetic field~\cite{sato.fujimoto.09,sato.takahashi.09,sato.takahashi.10}:
\begin{eqnarray}
\label{eq.relation} h_{c}^{2} = \left( 2 t - \mu \right)^{2} + | \Delta |^{2} .
\end{eqnarray}
With increase of magnetic field, the quasiparticle spectrum closes and reopens as a new topologically non--trivial gap at $h = h_{c}$~\cite{moore.balents.07}.
In the case of the dimerized SSH nanowire, emergence of non--trivial phase depends on the existence of additional parameters (e.g., $\lambda$ and $\delta$).
Then, value of $h_{c}$ depends on model parameters in non--trivial manner and can be determined analytically (more details can be found in Ref.~\cite{kobialka.sedlmayr.19}), but still, in the limit of $\delta \rightarrow 0$, condition~(\ref{eq.relation}) remains unchanged.

\paragraph*{Formalism. --}
The model Hamiltonian $\mathcal{H}$ can be numerically diagonalized by the Bogoliubov--Valatin transformation~\cite{degennes.89}:
\begin{eqnarray}
c_{is\sigma} &=& \sum_{n} \left( u_{isn\sigma} \gamma_{n} 
- \sigma v_{isn\bar{\sigma}}^{\ast} \gamma_{n}^{\dagger} \right) , 
\label{eq.bvtransform}
\end{eqnarray}
\AK{where $\gamma_{n}$, $\gamma_{n}^{\dagger}$ are the ``new'' quasiparticle fermionic operators.
This transformation yields Bogoliubov--de~Gennes equations in the form $\mathcal{E}_{n} \Psi_{isn} = \sum_{js'} \mathbb{H}_{is,js'} \Psi_{js'n}$, where $\mathbb{H}_{is,js'}$ is the Hamiltonian in the matrix form given as:
\begin{eqnarray}
\nonumber \mathbb{H}_{is,js'} &=& \left(
\begin{array}{cccc}
H_{is,js',\uparrow} & D_{is,js'} & S_{is,js'}^{\uparrow\downarrow} & 0 \\ 
D_{is,js'}^{\ast} & -H_{is,js',\downarrow}^{\ast} & 0 & S_{is,js'}^{\downarrow\uparrow} \\ 
S_{is,js'}^{\downarrow\uparrow} & 0 & H_{is,js',\downarrow} & D_{is,js'} \\ 
0 & S_{is,js'}^{\uparrow\downarrow} & D_{is,js'}^{\ast} & -H_{is,js',\uparrow}^{\ast}
\end{array} \right) \\
\label{eq.bdg}
\end{eqnarray}
while eigenvector:
\begin{eqnarray}
\Psi_{isn} &=& \left( u_{isn\uparrow} , v_{isn\downarrow} , u_{isn\downarrow} , v_{isn\uparrow} \right)^{T} .
\end{eqnarray}
The matrix block elements (taking into account both sublattices) are given here by $H_{is,js',\sigma} 
= - t (1+\delta) \delta_{ij} \delta_{\langle ss'\rangle} - t(1-\delta) \delta_{i-1,j} \delta_{\langle s,s'\rangle}
- ( \mu + \sigma h ) \delta_{ij} \delta_{ss'} $, the on-site superconducting gap is denoted by 
$D_{is,js'} = \Delta \delta_{ij} \delta_{ss'}$ and $S_{is,js'}^{\sigma\sigma'} = - i \lambda 
( \sigma_{y} )_{\sigma\sigma'} \left( (1+\delta) \delta_{ij} \delta_{\langle ss'\rangle} - ( 1- \delta ) \delta_{i-1,j} \delta_{\langle ss'\rangle} \right)$ 
stands for the spin-orbit Rashba term.
Here, we must keep in mind, that the indexes $i$ and $s$ change values over number of unit cells and sublattice indexes, respectively.
From this, $\mathbb{H}_{is,js'}$ is a square matrix with size of $4 \mathcal{N} \times 4 \mathcal{N}$, where $\mathcal{N}$ denotes number of sites in the system.
In the absence of the impurity $\mathcal{N}$ is equal to double of cells number $N_{\Omega}$.}

\AK{From solution of the BdG equations we can determine the spin resolved local density of states (LDOS) $\rho_{is \sigma} ( \omega ) 
=  -1/\pi \Imag \langle \langle c_{is\sigma} 
| c_{is\sigma}^{\dagger} \rangle\rangle$, which can be expressed 
by~\cite{matsui.sato.03}:
\begin{eqnarray}
\nonumber \rho_{is\sigma} ( \omega ) = \sum_{n} \left[ | u_{isn\sigma} |^{2} 
\delta \left( \omega - \mathcal{E}_{n} \right) + | v_{isn\sigma} |^{2} 
\delta \left( \omega + \mathcal{E}_{n} \right) \right] . \\
\label{eq.ldos}
\end{eqnarray}
Also spin polarization asymmetry (SPA) of the LDOS
\begin{eqnarray}
\delta \rho_{is} ( \omega ) =  \rho_{is \uparrow} ( \omega )  - \rho_{is \downarrow ( \omega )  }
\end{eqnarray}
can give additional information, e.g., about spin polarization of the bound state~\cite{sticlet.bena.12}.
\AK{In numerical calculations we replace the Dirac delta function by  Lorentzian $\delta ( \omega ) = \zeta / [ \pi ( \omega^{2} + \zeta^{2} ) ]$ with a small broadening  $\zeta / t = 0.001$.}
}

\AK{
Total LDOS $\rho_{is \uparrow} ( \omega )  + \rho_{is \downarrow} ( \omega )$ in low temperature limit gives information about the differential conductance $G(\omega)$~\cite{figgins.morr.10,chevallier.klinovaja.16,stenger.stanescu.17}.
Similarly, SPA LDOS $\delta \rho_{is}$ can give information about spin polarization of the bound states.
Both quantities can be measured  in relatively simple way using scanning tunneling microscope (STM)~\cite{hofer.foster.03,wiesendanger.09,oka.brovko.14}.
Experiments with magnetic tip  give information about magnetic structure of the bound states in atomic scale~\cite{meier.zhou.08,hus.zhang.17,rolfpissarczyk.yan.17}.
From the theoretical point of view, previous studies in spinfull models shown that the MBS has spin polarization
~\cite{sticlet.bena.12,maska.domanski.17,ptok.kobialka.17,kobialka.ptok.18.ring}.
From this, an existence of topological bound states can be probed via mentioned previously spin-polarized STM measurements~\cite{hofer.foster.03,wiesendanger.09,oka.brovko.14} (which has been done, e.g., in ferromagnetic atom chains~\cite{jeon.xie.17,kim.palaciomorales.18}).
This type of measurements can be useful in distinguishing between \AP{the ordinary Andreev bound states (ABS)} and topological MBS in hybrid nanostructures~\cite{devillard.chevallier.17}.
}

\AK{
Similar analysis of the system can be performed in the momentum space (more details can be found in Ref.~\cite{kobialka.sedlmayr.19}).
In this case, system can be studied via spin resolved spectral function $\mathcal{A}_{{\bm k}\sigma} (\omega) = - 1/\pi \ \Imag \langle \langle c_{{\bm k}\sigma} | c_{{\bm k}\sigma}^{\dagger} \rangle\rangle$.
From this, band structure and its SPA: $\delta \mathcal{A}_{\bm k} (\omega) = \mathcal{A}_{{\bm k}\uparrow} (\omega) -\mathcal{A}_{{\bm k}\downarrow} (\omega)$ can be found~\cite{bansil.lin.16}.
Similarly to LDOS, these quantities can be measured via the angle-resolved photoemission spectroscopy (ARPES) technique~\cite{damascelli.zahid.03}, even in nanostructures~\cite{snijders.weitering.10}.
Existence of the topological phase in the system leads to the observation of the band inversion, clearly visible in the spin polarization of bands.
This is typical not only for the case of the topological insulator~\cite{hasan.kane.10,bansil.lin.16}, but also for other systems in which the topological phase emerges~\cite{szumniak.chevallier.17,kobialka.ptok.18.ring,sticlet.pascumoca.20}.}

\section{Numerical results}
\label{sec.results}

In this section we will discuss the leakage of MBS to the impurity within the dimerized SSH nanowire.

As for the parameters used in calculation, \AK{we took a nanowire composed of $N_{\Omega} = 100$ cells, i.e. $\mathcal{N}=200$ sites and an additional impurity being a $201^{st}$ site (unless stated otherwise).}
\AK{Alternating order of bonds is preserved in the junction between nanowire and impurity.}
Nanowire is characterized by  $\Delta / t = 0.2$ and $\lambda / t = 0.15$.
\AK{Any change in chemical potential $\mu$ affects whole system, both the nanowire and impurity.}
Here, it should be mentioned, that the described results do not depend on the size of the nanowire. 
Additionally, \AK{throughout whole paper,} we take $h = 0.3 t> h_{c}$, which ensures that the homogeneous system is in the non--trivial phase.
If not stated differently, when nanowire has odd number of sites, it begins with a weak $(1-\delta)t$ bond and ends with a strong $(1+\delta)t$ bond.

\AK{The existence of hopping modulation has a negative impact on the usual non--trivial phase.}
However, for the dimerization--dependent branch it is essential for its existence.
Let us start with discussion of the influence of the impurity on Rashba nanowire.

\begin{figure}[!t]
\includegraphics[width=\linewidth,keepaspectratio]{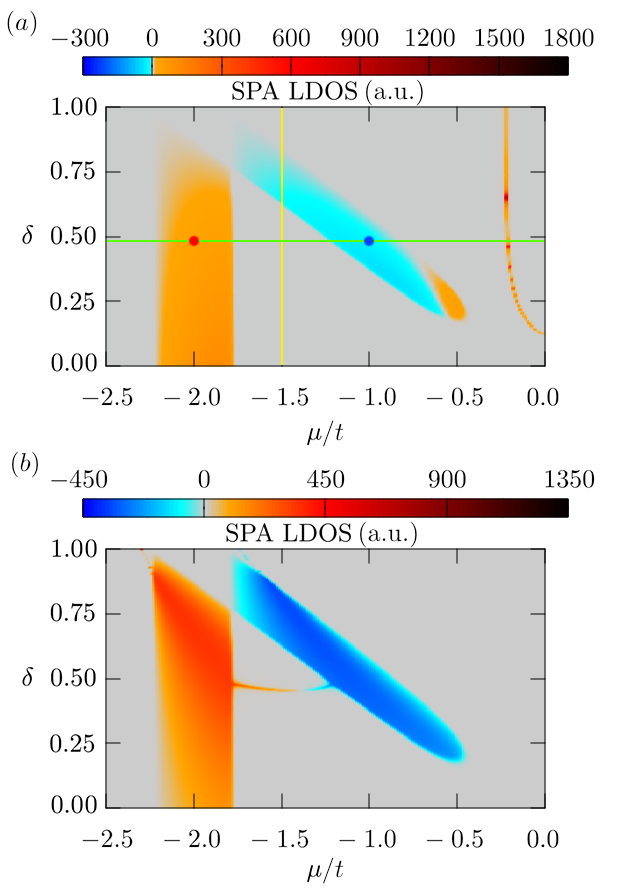}
\caption{
\AK{SPA LDOS for zero energy $\mu$--$\delta$ phase space of first  site (a) of the system and impurity site (b).
First site is connected to the rest of the nanowire with a weak bond, while impurity site is connected by strong bond.
In the case of (b) the impurity is connected with strong bond, which allows for forming of  the {\it bridge}-like structure.
Eigenvalues for parameters along green and yellow lines at (a) are shown in Fig.~\ref{fig.mudeltaeig}.
Red and blue dots correspond to $\delta \mathcal{A}_{\bm k}(\omega)$ for different topological phases, presented on Fig.~\ref{fig.pasma}.
Results for system with $\mathcal{N} = 201$ sites and $h/t=0.3$.
}
}
\label{fig.mudelta}
\end{figure}

\AK{In the Fig.~\ref{fig.mudelta}, we show color map of SPA LDOS for $\omega = 0$, as a function of chemical potential $\mu$ and hopping modulation $\delta$.
Nanowire is in the presence of the impurity, which is connected to the system with strong bond $(1+\delta)$.}
In Fig.~\ref{fig.mudelta}, regions centered around $\mu = 2t$ (near the bottom of the band) show parameters of system which allow for hosting of MBS in system.
This range of parameters, where the non-trivial phase exist, can be associated with typical limit in the homogeneous system~\cite{sato.takahashi.09,sato.fujimoto.09,sato.takahashi.10}.
Additional modulation of hopping introduced by $\delta$ does not change the topological character of the system in $\delta \rightarrow 0$.
However, 
bond modulation creates additional topological branch that allows for existence of MBS in broader range of parameters, in accordance to Eq.~(\ref{eq.relation}).
This additional, dimerized branch incorporates regions within the band that for some range of modulation of hopping integral a non--trivial phase appears in which MBS can emerge.
The abrupt change of the SPA of the system between two branches of topological phase can be explained by the reordering of bands that happens with each band closure at the instance of topological transition~\cite{kobialka.sedlmayr.19}.
When the bands close at the transition from topologically non--trivial to trivial state ($\mu \simeq 1.8t$), they reopen in opposite order during the transition to non--trivial state (within a dimerized topological branch)~\footnote{cf.~Supplemental Material in Ref.~\cite{kobialka.sedlmayr.19}}.
Due to the interplay between magnetic field and SOC the change of spin polarization occurs. 
In Fig.~\ref{fig.mudelta}(a) we can see a phase space for the first site of nanowire, linked to main part of nanowire with weak bond $(1-\delta)$. 
This allows for visualization of characteristic feature for investigated system, a parabola at $\mu/t \in ( -0.22,0.22 )$ (as plots are $\mu$--symmetric), which is a manifestation \AK{of  states of first site of the nanowire}, crossing at zero energy.
Another distinctive feature can be seen on panel Fig.~\ref{fig.mudelta}(b), where we can see a \AK{SPA LDOS space for the impurity} (being a last site of nanowire) which is linked to main part of nanowire with strong bond $(1+\delta)$.

\begin{figure}[!t]
\includegraphics[width=\linewidth,keepaspectratio]{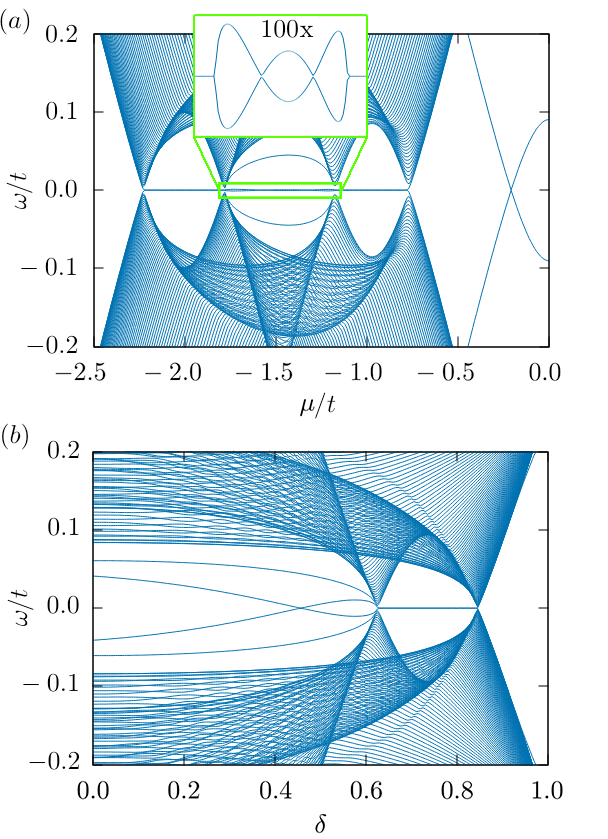}
\caption{
\AK{
Spectrum of the system for fixed $\delta=0.46$ (a) and $\mu/t=-1.5$ (b), which corresponds to green horizontal and yellow vertical  lines in Fig.~\ref{fig.mudelta}(a), respectively.
Results in absence of impurity are presented at Fig. 8 in Ref.~\cite{kobialka.sedlmayr.19}.
}
}
\label{fig.mudeltaeig}
\end{figure}

\begin{figure}[t]
\includegraphics[width=0.9\linewidth,keepaspectratio]{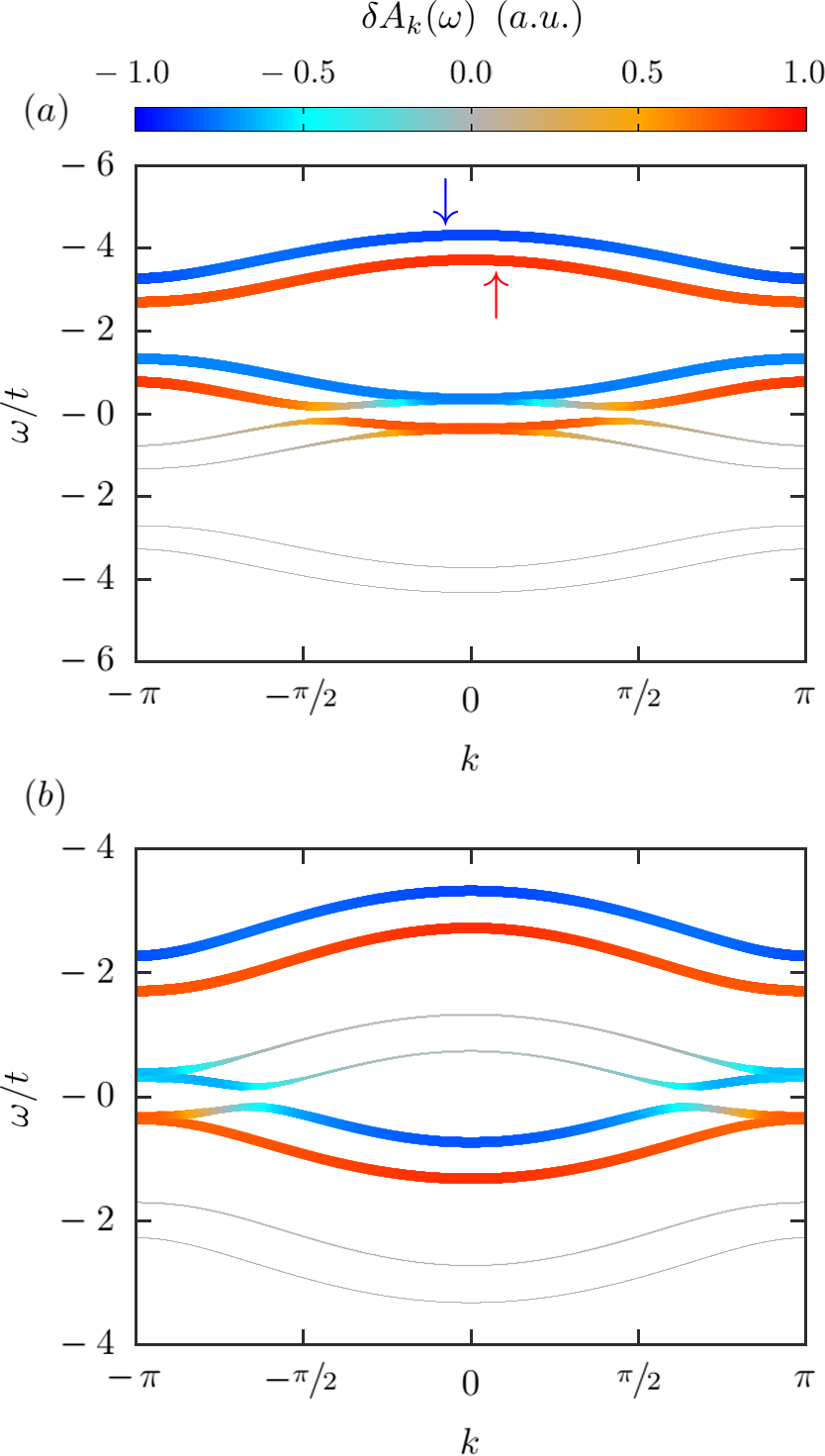}
\caption{
\AK{The SPA of the spectral function $\delta \mathcal{A}_{\bm k} (\omega)$.
Results for (a) $\mu/t = -2$ and (b) $\mu/t = -1$, with fixed $\delta  = 0.46$ (cf. red and blue dots in Fig.~\ref{fig.mudelta}(a), respectively).
Color corresponds to spin polarization (marked with corresponding arrows) and width of line to total spectral function $\mathcal{A}_{{\bm k}\uparrow}(\omega)+\mathcal{A}_{{\bm k}\downarrow}(\omega)$. }
}
\label{fig.pasma}
\end{figure}

\begin{figure}[!t]
\includegraphics[width=\linewidth,keepaspectratio]{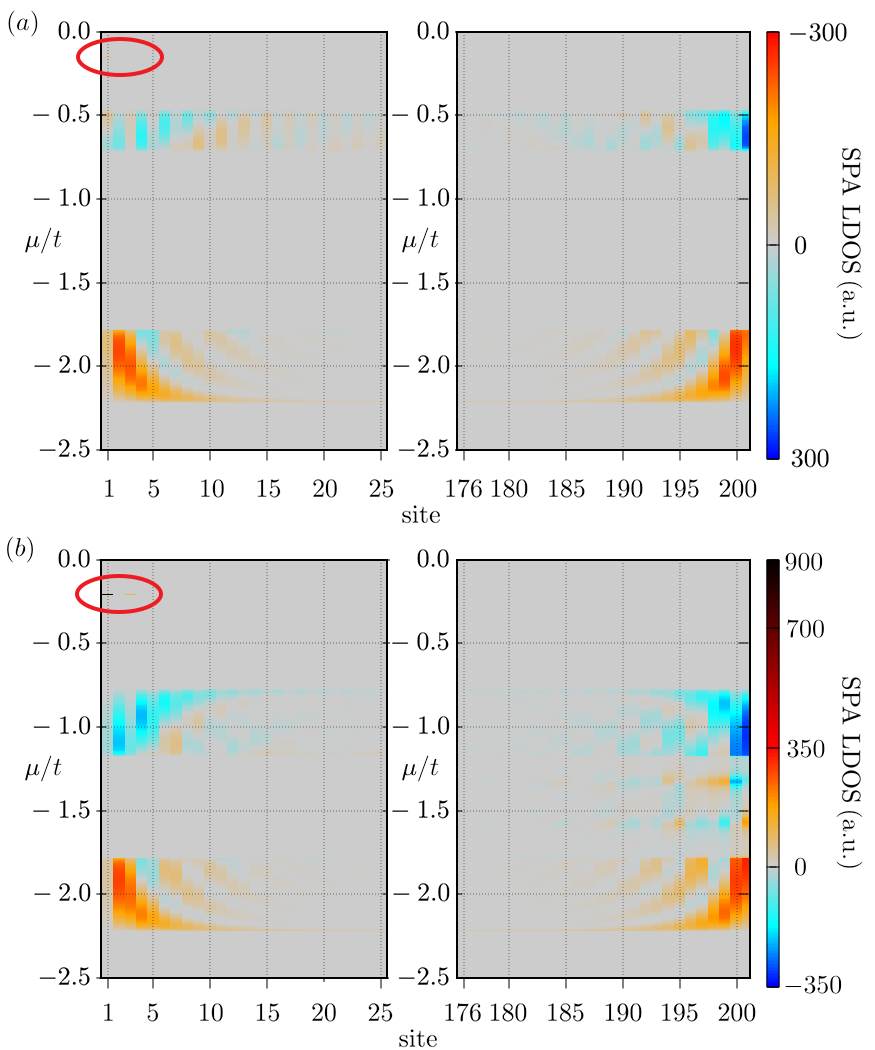}
\caption{Real space distribution of asymmetry of SPA of MBS as a function of chemical potential, at the ends of the nanowire.
Results for parameters like in Fig.~\ref{fig.mudelta} along $\delta=0.24$ (a) and $\delta=0.46$ (b).
Central feature for $\mu/t \simeq -1.5$ shows the distribution of SPA LDOS along the {\it bridge}-like structure from Fig.~\ref{fig.mudelta}.
\AK{Regions within red oval show instances of weak bond parabola state forming on first site. }
\AK{Here, our system consists of $\mathcal{N} = 200$ sites.}}
\label{fig.qd}
\end{figure}

Existence of the \textit{bridge}-like features can be understood from analysis of the system spectrum presented at Fig.~\ref{fig.mudeltaeig}.
There, a \textit{bridge}-like feature emerges due to the existence of ABS, connecting separate topological phases.
It is a result of crossing the Fermi level by the eigenvalues of states associated with the existence of impurity, coupled to nanowire by a strong bond.
There is no analogue of strong bond feature for a last site when it is not an impurity.
In Fig.~\ref{fig.mudeltaeig}(a) we can see eigenvalues for $\delta = 0.46$ [green line at Fig.~\ref{fig.mudelta}(a)], crossing the \textit{bridge}-like structure.
Here, the two zero energy Majorana states are separated by trivial \textit{bow tie}-like ABS feature (inset).
This in-gap state are also clearly visible in the SPA LDOS analyses and are strongly associated with the localization of ABS from one site of the nanowire -- near the impurity, cf.~Fig.~\ref{fig.qd}.
As we can see, this structure is in fact a manifestation of zero energy crossing of ABS.
Similar behavior can be observed in the case of the spectrum of the system from $\mu/t = -1.5$ [yellow line at Fig.~\ref{fig.mudelta}(a)], shown at Fig.~\ref{fig.mudeltaeig}(b).

\AK{
In contrast to the MBS in isotropic chain ($\delta=0$), in our results SPA LDOS of MBS has opposite value in different part of phase space (Fig.~\ref{fig.mudelta}).
This behavior is strongly associated with influence of $\delta$ into band structure and its spin polarization (Fig.~\ref{fig.pasma}).
Exact analysis of the band structure where MBS exist~\cite{kobialka.ptok.18.ring}, shows that the Majorana quasiparticle {\it inherits} spin polarization of bands nearest the zero energy, i.e. Fermi level.
Here, from studying the band structure, we can observe that the MBS in main branch has typical spin polarization [Fig.~\ref{fig.pasma}(a)], i.e. $\uparrow$.
In this case, emergence of the topological phase is associated with band inversion around $k=0$.
On contrary, SPA of the MBS in dimerization-dependent branch is opposite, i.e. $\downarrow$.
This is a consequence of the band inversion of the nearly fully filled bands around $k=\pi$ point [Fig.~\ref{fig.pasma}(b)].
Summarizing, in our case the SPA LDOS result yield unexpected results if compared to aforementioned results.
}

Crossing point shows accidental nature of \textit{bridge}-like feature of zero energy ABS.
Thanks to this, it is certain that region connecting two topological branches does not hold MBS, as this would result not only in zero energy state typical for MBS but additionally with an avoided crossing of ABS.
\AK{On the other hand, if the nanowire would be pristine (no impurity), near--zero energy states that do not mutate into MBS after the topological transition would not show any avoided crossing or \textit{bow tie} behavior, but instead, follow the MBS--would be state and diverge out of the topological regime.}

Now, we discuss the zero--energy SPA LDOS, shown at Fig.~\ref{fig.qd}.
In the case of non--trivial phase, the MBS are localized at both ends of nanowire.
These states are characterized by the oscillation of the SPA LDOS in space.
As we can see, in both branches of the non--trivial phase, LDOS is characterized by opposite SPA. 
Largest localization of the state is visible at the impurity site (right hand side) -- cf. $\mu/t \sim -2$ main branch and $\mu/t \sim -0.75$ dimerized branch, \AK{ABS are pinned to the impurity.}
For the intermediate region ($\mu/t \sim -1.5$) we observe localization of the state mostly at impurity, what is associated with  ABS states mentioned before manifested as \textit{bridge}-like structure in the phase space \AK{and, correspondingly, a \textit{bow tie} region in eigenvalues of~Fig.~\ref{fig.mudeltaeig}.
As we move away from the impurity towards the middle of the nanowire, \textit{bridge}--like feature will fade away and show no SPA within a distance of about $\sim20$ sites.}
\AK{Additionally, we can observe instances of weak bond parabola states forming on first site (red ovals).
These states are characterized by high SPA LDOS and correspond to ABS forming on edge site which is weakly connected to the rest of nanowire.}

\begin{figure}[!t]
\includegraphics[width=\linewidth,keepaspectratio]{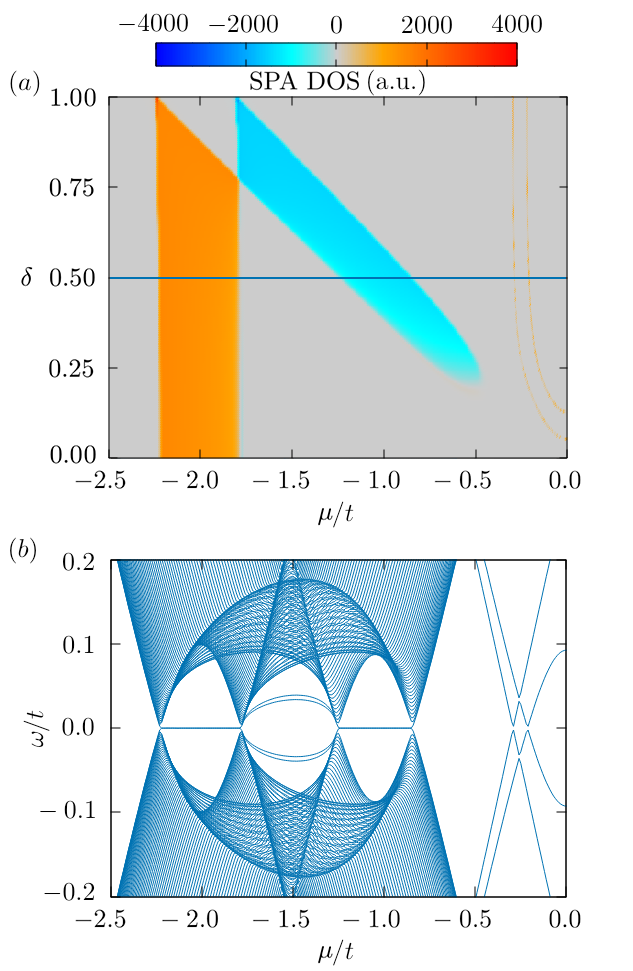}
\caption{
(a) SPA for zero energy DOS phase spaces as a function of $\mu$ and $\delta$, in the case of when nanowire begins with weak bond and connects with a weak bond to impurity (cf.~Fig.~\ref{fig.mudelta}).
\AK{Here, our system consists of $\mathcal{N} = 200$ sites.}
(b) Eigenvalues for the cross-section of (a), along the blue line ($\delta=0.5$, cf.~Fig.~\ref{fig.mudeltaeig}).
}
\label{fig.200}
\end{figure}

\AK{
Here, we should also discuss an important problem of interplay between trivial energy levels (of quantum dot or impurity) with energy levels of SSHR chain, which in topological regime contains MBS.
In typical case, when additional impurity is connecting to trivial superconducting system, the ordinary in-gap Andreev bound states emerge~\cite{balatsky.vekhter.06}.
Situation is more interesting when impurity is connected to the superconducting system in topological phase.
For instant, this issue was experimentally studied by Deng {\it et al.}~\cite{deng.vaitiekenas.17}, in fabricated nanowire with a quantum dot at one end.
Topologically trivial bound states were seen to coalesce into MBSs as the magnetic field was increased.
Theoretical study of this behavior showed, that the interplay between trivial ABS and topological MBS strongly depends by spin polarization of the ABS~\cite{prada.aguado.17,ptok.kobialka.17,szumniak.chevallier.17,hoffman.chevallier.17}, due to positive spin polarization of the MBS~\cite{kobialka.ptok.18.ring}.
In such case, the avoided crossing or resonance of the ABS energy levels can be observed~\cite{vernek.penteado.14,baranski.kobialka.16,zienkiewicz.baranski.19}.
Moreover, this behavior can be helpful in distinguishing MBS from ABS~\cite{liu.sau.18,ricco.desouza.19}.
Here we must have in mind, that the boundary of the topological regime of one dimensional nanowire is given by relation~(\ref{eq.relation})~\cite{zhang.nori.16,szumniak.chevallier.17}.
In this regime, MBS has the same spin polarization~\cite{kobialka.ptok.18.ring}.
Contrary to this, in discussed SSHR model, topological phase diagram has more complicated form --- due to the existence of the main and dimerization-dependent branches (cf.~Fig.~\ref{fig.mudeltaeig})~\cite{kobialka.sedlmayr.19}.
Here, spin polarization of the MBS depends on parameters of the system, i.e. in main (dimerized) branch it is positive (negative).
Unfortunately, this can lead to ambiguity in distinguishing between ABS and MBS states.
}

Now, we analyze results for \AK{system with even ($200$ in total)} number of sites.
In such a case nanowire begins with weak bond and connects with a weak bond to impurity (Fig.~\ref{fig.200}).
Number of sites does not affect the results in any other way than just the order of weak/strong bonds.
Here, we can see a familiar phase space with two additional parabolas in zero--energy SPA LDOS [Fig.~\ref{fig.200}(a)] forming at $\mu < 0.3 t$.
$\mu$--position of starting point for outer parabola is linearly dependent on the value of magnetic field.
As for the inner parabola, its forms only if the system exists in non--trivial phase, after the gap closing ($h > h_c$), similar to the \textit{bridge}-like feature.
If nanowire would start and end with a strong bond, a \textit{bridge}-like feature identical to the one from Fig.~\ref{fig.mudelta}(b) would appear. 
However, the fact of both bonds being the same would not affect the \textit{bridge} in any way, in contrast to the situation with weak bond.

\section{Summary}
\label{sec.sum}

In this paper we have shown that Majorana bound state leakage in Rashba nanowire that is dimerized according to the SSH scenario might behave anomalously when additional impurity is in the vicinity of nanowire.
We find that topological branches, usual and dimerized one, have different SPA that can be explained by opposite order of bands taking part in topological transitions, \AK{which are closest to Fermi level}. 
Moreover, introduction of impurity along the dimerized nanowire influences the leakage profile of Majorana state into the trivial impurity.
Coupling of impurity to nanowire leads to emergence the trivial Andreev bound states, strongly localized around the impurity.
In the case of the one-site impurity, this can lead to emergence of states crossing the Fermi level.
In consequence we observe trivial zero--energy states in form {\it bridge}-like structure, connecting two branches of the non--trivial topological phases.
\AK{Stemming from this, measurements of both ends of nanowire in search of MBS could resolve an ambiguity created by potential existence of impurities in nanowire.}

\begin{acknowledgments}
We thank Pascal Simon and Nicholas Sedlmayr for inspiriting discussions.
This work was supported by the National Science Centre (NCN, Poland) under grants 
UMO-2031/N/ST3/01746 (A.K.), 
and
UMO-2017/25/B/ST3/02586 (A.P.). 
\end{acknowledgments}

\bibliography{biblio}

\end{document}